\documentclass[runningheads]{llncs}

\usepackage[T1]{fontenc}
\usepackage{graphicx}

\usepackage{amsmath}
\usepackage{adjustbox}
\usepackage{siunitx}
\usepackage{csquotes}

\usepackage{url}

\newrobustcmd\B{\DeclareFontSeriesDefault[rm]{bf}{b}\bfseries}

\newcolumntype{H}{>{\setbox0=\hbox\bgroup}c<{\egroup}@{}}

\begin{document}
\title{Detection of Prosodic Boundaries in Speech Using Wav2Vec 2.0}
%
%
\author{Marie Kune\v{s}ov\'a\orcidID{0000-0002-7187-8481} \and
Mark\'eta \v{R}ez\'a\v{c}kov\'a\orcidID{0000-0002-6194-7826}
}

\authorrunning{M. Kune\v{s}ov\'a and M. \v{R}ez\'a\v{c}kov\'a}

\institute{New Technologies for the Information Society and Department of Cybernetics,\\
Faculty of Applied Sciences, University of West Bohemia, Pilsen, Czech Republic\\
\email{\{mkunes,juzova\}@ntis.zcu.cz}
}

\maketitle              
\begin{abstract}

Prosodic boundaries in speech are of great relevance to both speech synthesis and audio annotation. In this paper, we apply the wav2vec 2.0 framework to the task of detecting these boundaries in speech signal, using only acoustic information. We test the approach on a set of recordings of Czech broadcast news, labeled by phonetic experts, and compare it to an existing text-based predictor, which uses the transcripts of the same data. Despite using a relatively small amount of labeled data, the wav2vec2 model achieves an accuracy of 94\% and F1 measure of 83\% on within-sentence prosodic boundaries (or 95\% and 89\% on all prosodic boundaries), outperforming the text-based approach.
However, by combining the outputs of the two different models we can improve the results even further. 

\end{abstract}

\footnotetext[0]{This preprint is a pre-review version of the paper and does not contain any post-submission improvements or corrections. The Version of Record of this contribution is published in the proceedings of the International Conference on Text, Speech, and Dialogue (TSD 2022), LNAI volume 13502, and is available online at \url{https://doi.org/10.1007/978-3-031-16270-1_31}}

\section{Introduction}

Prosodic phrasing is the division of fluent speech into \emph{prosodic} (or \emph{intonation}~\cite{beckman1997guidelines}) \emph{phrases} -- groups of words in a spoken sentence, typically featuring an intonation peak and often separated by pauses.

Prosodic phrasing not only plays an important role in the human understanding of spoken language~\cite{frazier2006prosodic} but is also highly relevant for many speech processing tasks, such as speech synthesis and audio annotation. 

In text-to-speech (TTS) systems, information about prosodic boundaries in text helps improve the naturalness of synthesized speech, by allowing the system to insert pauses and modify intonation in a similar way to a human speaker. In audio, it can be used to enhance the training data, likewise leading to a more natural-sounding speech~\cite{Taylor_2009}.

In speech recognition and spoken language understanding, phrase breaks also help distinguish between otherwise identical sentences with a different meaning (such as the popular example \emph{\enquote{Let's eat, grandma!}} versus \emph{\enquote{Let's eat grandma!}}).

There are two different scenarios for the automatic detection of prosodic boundaries: detection solely from text, most often for the purposes of speech synthesis~\cite{futamata21_interspeech,read2007stochastic,taylor1998assigning,volin2021human,zou21_interspeech}, or detection from spoken utterances as a form of audio annotation. In the latter case, some approaches have been based solely on acoustic information (though sometimes with word or syllable boundaries derived from text transcripts)~\cite{lin20m_interspeech,rosenberg2010autobi,Schuppler2020,suni2016boundary}, while others have combined both lexical and acoustic information~\cite{Christodoulides2017,gallwitz2002integrated,KOCHAROV2019231}.

In this paper, our main goal is to obtain a detector which works solely in the audio modality, using only acoustic cues. However, its results will also be compared to an existing text-based model~\cite{volin2021human}, evaluated on the transcripts of the same utterances.

\section{Data}

The experiments were performed on a set of recordings of Czech radio broadcast news (Channels 1 and 2 of the Czech Radio), previously used in \cite{volin2021human} as the News-Reading Speech (NRS) corpus\footnotemark. The dataset consists of 12 news bulletins presented by different speakers (six male and six female), each between 2.5 and 5 minutes long, for a total of 42 minutes of speech (486 sentences). The recordings have been annotated by phonetic experts, following the guidelines in \cite{beckman1997guidelines}. 

\footnotetext{Since the publication of \cite{volin2021human}, the NRS annotations have undergone a round of revisions and the model was updated accordingly. The text-based results in section~\ref{sec:textresults} will thus differ from those listed in the aforementioned paper.}

The annotation conventions, as described in \cite{beckman1997guidelines}, include multiple levels of phrasing: most relevantly, prosodic (intonation) phrases can also be further divided into one of more \emph{intermediate} phrases -- smaller units with less discernible boundaries. These are also labeled in the NRS dataset. However, in our work, we are specifically interested in the detection of \emph{prosodic} boundaries as the most important ones for most speech processing applications -- we will explore the use of intermediate boundaries during \emph{training}, but we ignore them during evaluation. 

\section{Model for text-based detection}

We compare the results of our audio-based prosodic boundary detection to those of our existing text-based detector~\cite{volin2021human}, which was tested on the same dataset. 

This model remains as described in~\cite{volin2021human}: it is a Text-to-Text Transfer Transformer (T5) model~\cite{raffel2020exploring}, which transforms a given sequence of words into an output sequence with predicted phrase boundaries. It was pre-trained on large amounts of unlabeled Czech text in the CommonCrawl corpus and fine-tuned for the phrase detection task on what~\cite{volin2021human} referred to as The Laboratory Speech (LS) data -- text sentences from 6 large-scale Czech speech corpora created for the purposes of speech synthesis in the TTS system ARTIC \cite{tihelka2018current}. 

The prosodic boundaries in the LS dataset were labeled only using automatic segmentation, but the fine-tuned model was subsequently adapted on the hand-annotated NRS data using a leave-one-out approach -- 12 different models were trained, each adapted on 11 speakers and evaluated on the last speaker.

\section{Model for audio-based detection}

Systems for audio-based prosodic boundary detection have traditionally utilized combinations of different features such as the duration of pauses and syllables, $F_0$ range and resets, intensity, or pitch movement~\cite{Christodoulides2017,KOCHAROV2019231,lin20m_interspeech,rosenberg2010autobi,Schuppler2020}. Rather than use such handcrafted combinations of features, however, we chose to employ learned representations from raw audio data.

Wav2vec 2.0~\cite{baevski2020wav2vec} is a self-supervised framework for speech representation which has been used for a large variety of different speech-related tasks~\cite{cooper2021generalization,yang21c_interspeech,zhang2020pushing}. One of the main advantages of the wav2vec approach is that a generic pre-trained model can be fine-tuned for a specific purpose using only a small amount of labeled data.

We use the pre-trained wav2vec 2.0 base model \enquote{ClTRUS}\footnote{\textbf{C}zech \textbf{l}anguage \textbf{TR}ransformer from \textbf{U}nlabeled \textbf{S}peech, \newline available from: \url{https://huggingface.co/fav-kky/wav2vec2-base-cs-80k-ClTRUS}}, which is specifically trained for the Czech language using more than 80 thousand hours of Czech speech from various domains~\cite{lehecka2022cltrus}.

\begin{figure}
    \centering
    \includegraphics[width=\textwidth]{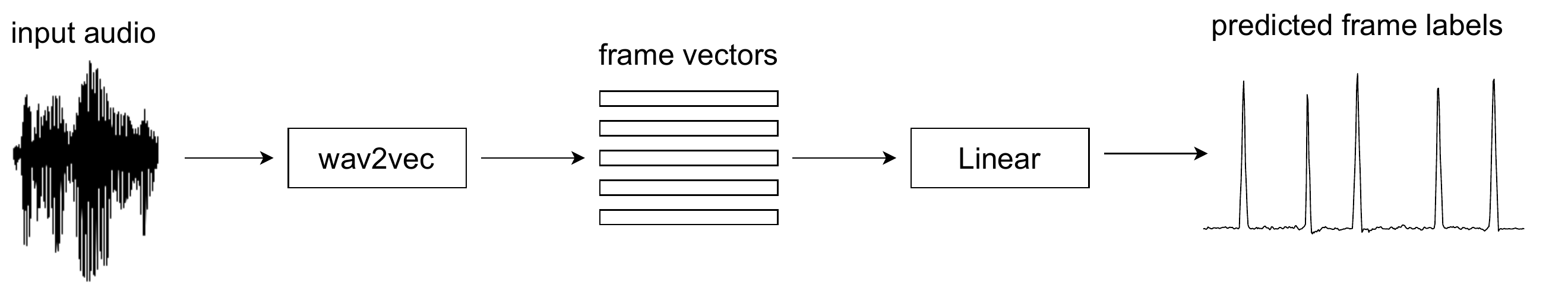}
    \caption{Illustration of the wav2vec2-based prosodic boundary detector. The model outputs a label for each audio frame (every 20\,ms).}
    \label{fig:w2v_diagram}
\end{figure}

Using the HuggingFace Transformers library~\cite{wolf2020transformers}, we fine-tuned the model for an audio frame classification task (\emph{Wav2Vec2ForAudioFrameClassification}) on the NRS data (Fig.~\ref{fig:w2v_diagram}) and evaluate it using a leave-one-out approach, similarly to the text-based T5 model.

\begin{figure}
    \centering
    \adjincludegraphics[width=\textwidth,trim={{.1\width} 0 {.08\width} 0},clip]{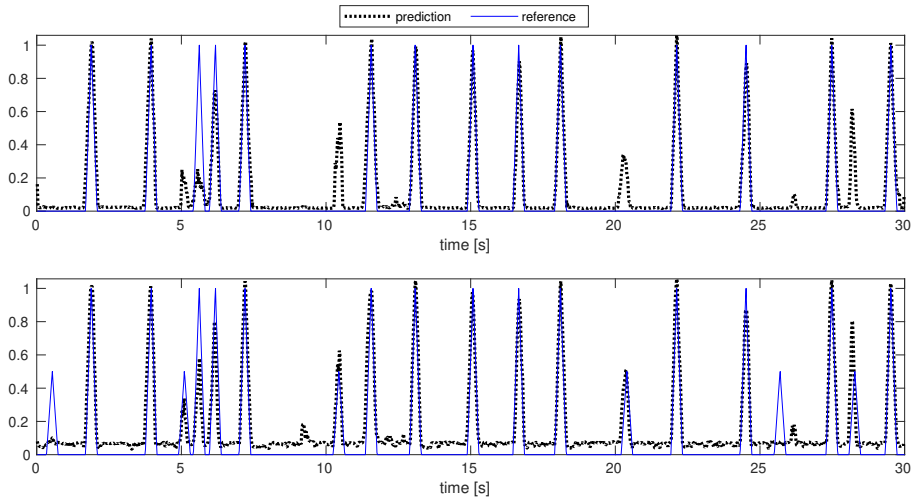}
    \caption{Example of the reference labels and predictions for one audio segment. Training labels for the wav2vec2 model either include only prosodic phrase boundaries, with a peak value of 1 (top), or also intermediate phrase boundaries, with a smaller peak value of 0.5 (bottom).}
    \label{fig:w2v_ref_labels}
\end{figure}

During the fine-tuning of the wav2vec 2.0 model, the references are given in the form of a fuzzy labeling function, as depicted in Figure~\ref{fig:w2v_ref_labels} (top): prosodic boundaries are given the reference label 1, linearly decreasing to 0 in an interval $\pm 0.2$\,s around each boundary. 

The model was fine-tuned with MSE loss. The fine-tuning process is very fast -- the model learns to predict the triangular shapes nearly perfectly within several epochs, at which point the results do not improve further with additional training.

Due to the relatively high memory requirements of wav2vec, the audio is processed in chunks of 30\,s, with a 15\,s step -- the chunks are partially overlapping. When the outputs are stitched back together for evaluation, the middle part of each chunk is used and the overlapping edges are discarded. This was originally meant to avoid potential issues near the beginning and end of each chunk (due to missing context on one side). However, in terms of the overall precision and recall, the difference appears to be minimal.

Finally, in order to improve the robustness and consistency of the results and limit the influence of random chance, each model was fine-tuned five times with identical settings and different random seeds, and the raw outputs were averaged. This does not substantially improve the results, but it reduces random fluctuations and allows for a better comparison between models fine-tuned with different settings. 

\subsection{Influence of intermediate phrase boundaries}

As previously stated, our targets for prediction are only the prosodic phrase boundaries. However, the less important intermediate boundaries may also convey useful information for training, particularly since the distinction between the two categories is not always clear.

In our initial experiments with a model fine-tuned solely on prosodic boundaries, we found that the majority of false positives (approximately two thirds, as seen in Figure~\ref{fig:prec_vs_rec_audio}) were located in spots marked as \emph{intermediate} phrase boundaries by the expert annotators. This is despite the fact that intermediate boundaries are present in less than 7\% of all word boundaries in the NRS dataset.

This spurred us to question whether it is truly appropriate to label these boundaries as \emph{zero} in the reference labels - they clearly exhibit similar acoustic features that the model is learning to detect, albeit perhaps to a less pronounced degree. Assigning them a smaller, but non-zero label may have a positive effect on the resulting model.

Thus, we decided to test two options for the training data:
\begin{enumerate}
    \item[a)] Only prosodic phrases are included in the reference labels.
    
    \item[b)] Both prosodic and intermediate phrases are included in the reference labels, with different values. Prosodic boundaries are given the maximum value of 1 and intermediate boundaries are labeled as 0.5 -- both with a linear decrease to zero over $\pm0.2$\,s, as previously described.
\end{enumerate}

In both cases, the model is still evaluated on \emph{prosodic} boundaries only.

\subsection{Post-processing}
\label{sec:postprocessing}

The wav2vec2 model outputs predicted labels for each audio frame (every 20\,ms). However, the text-based T5 model naturally predicts phrase breaks between words and is evaluated in terms of within-sentence word boundaries. Thus, it is necessary to convert the wav2vec2 predictions to a more comparable format:

First, we identify the peaks in the raw outputs. If the value of a peak is higher than a specific threshold and there is no higher peak within 0.25\,s, the system marks this as a predicted boundary. 

For the purposes of evaluation, these predicted boundaries are then aligned to the nearest end of a word within 100 ms, based on the reference annotations -- this is because the ground truth phrase boundaries are likewise aligned to the ends of words. 

For the numeric results listed in this paper, we did not specifically tune the decision threshold -- we simply choose the value 0.5 as the \enquote{middle ground}. Similarly, for the model trained with added intermediate boundaries, the threshold was selected as 0.75 -- as the average between the labels of prosodic boundaries (1.0) and intermediate boundaries (0.5). 

\section{Results}

In this paper, we list two separate sets of results, evaluated under slightly different conditions: First, evaluation of the full outputs of the wav2vec2 models, given as time labels, and including all boundaries, even those between sentences.

However, for a fair comparison with the T5 model, we secondly convert our predictions into text form (using the transcripts to ensure identical sentences), with boundaries marked only between words and ignoring the ends of sentences -- this is because the text-based T5 model worked with isolated sentences and only searched for prosodic boundaries \emph{within} the sentence.

\subsection{Evaluation measures}

The standard evaluation metrics for phrase boundary detection are precision ($P$), recall ($R$), accuracy (Acc), and F1-score, given as
\begin{equation}
\label{eq:prec}
    P = \frac{tp}{tp + fp}\\
\end{equation}
\begin{equation}
    R = \frac{tp}{tp + fn}\\
\end{equation}
\begin{equation}
    \operatorname{Acc} = \frac{tp + tn}{tp + tn + fp + fn}\\
\end{equation}
\begin{equation}
\label{eq:F1}
    F1 = 2 \cdot \frac{P \cdot R}{P + R}\\
\end{equation}
where $tp$ refers to the number of correctly detected phrase boundaries (\emph{true positives}), $fp$ the number of \emph{false positives}, $fn$ is the number of missed phrase boundaries (\emph{false negatives}) and $tn$ is the number of \emph{true negatives} - between-word boundaries that were correctly labeled as not being phrase breaks.

As the wav2vec2 model outputs per-frame predictions, not constrained to word boundaries, we decided to also perform frame-wise evaluation, in terms of segmentation of each audio file into prosodic phrases. For this, we chose segment purity and coverage (e.g. \cite{Bredin2017}) as the main metrics. These are obtained as
\begin{equation}
	\operatorname{purity}(S, R) = \frac{\sum_k \max_j |s_k \cap r_j|}{\sum_k |s_k|}
	\label{eq:purityBredin}
\end{equation} 
and
\begin{equation}
\operatorname{coverage}(S, R) = \frac{\sum_j \max_k |s_k \cap r_j|}{\sum_j |r_j|}
\label{eq:coverageBredin}
\end{equation} 
where $S = \{s_1, \ldots, s_K\}$ is the set of segments (i.e. prosodic phrases) found by the system, $R = \{r_1, \ldots, r_J\}$ corresponds to the reference segments, $|r_j|$ is the duration of segment $r_j$, and $s_k \cap r_j$ denotes the intersection of segments $s_k$ and $r_j$.

\subsection{Audio-based evaluation}
\label{sec:audioresults}

\begin{figure}
    \centering

    \adjincludegraphics[width=0.49\textwidth,trim={{.02\width} 0 {.08\width} 0},clip]{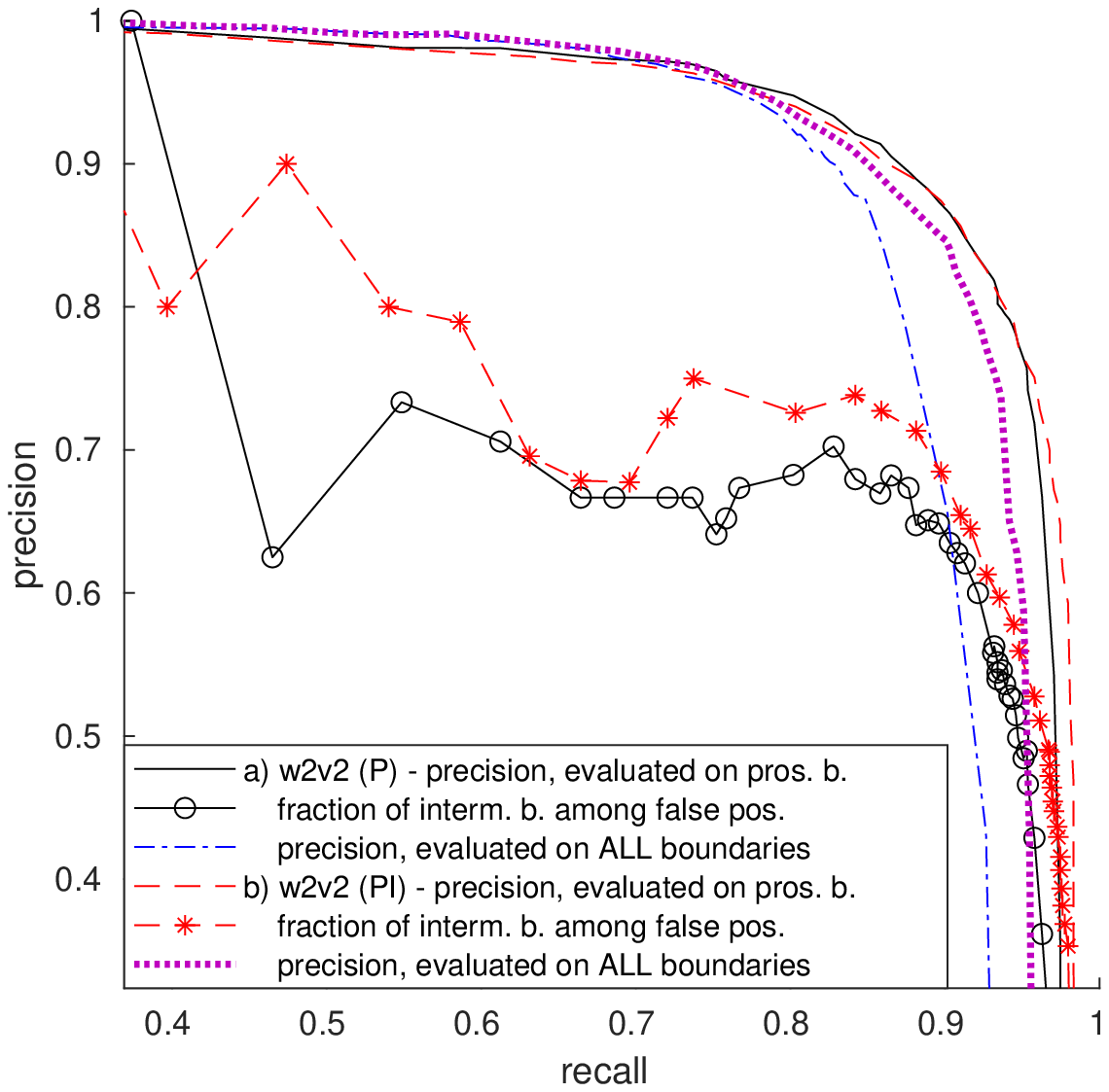}
    \adjincludegraphics[width=0.49\textwidth,trim={0 0 {.08\width} 0},clip]{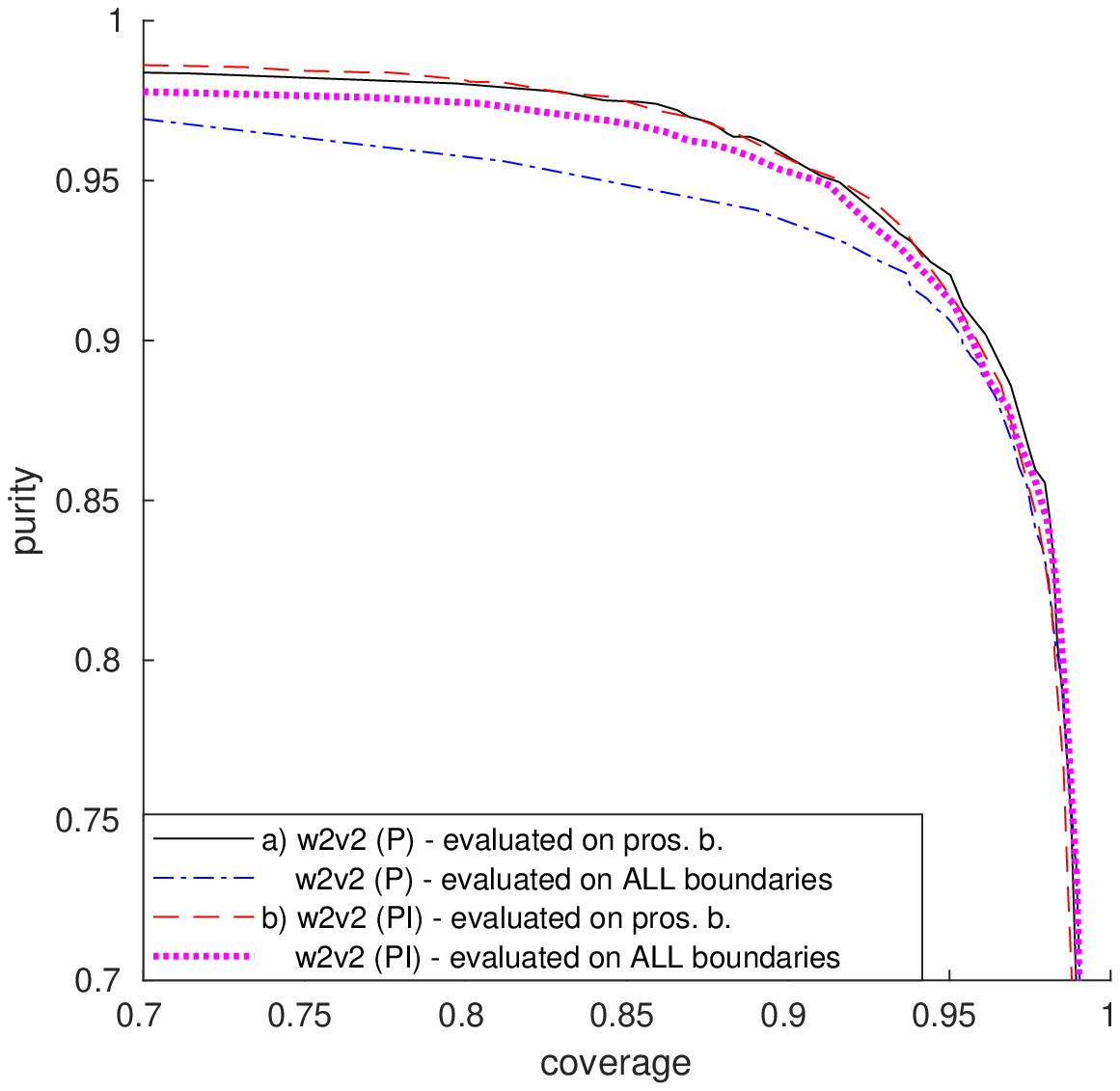}
    \caption{Precision-recall (left) and purity-coverage (right) curves of models fine-tuned a) only on prosodic boundaries (\enquote{w2v2 (P)}), or b) on both prosodic and intermediate boundaries (\enquote{w2v2 (PI)}). The left plot additionally shows the fraction of false positives which correspond to intermediate boundaries, relative to the total number of false positives.}
    \label{fig:prec_vs_rec_audio}
\end{figure}

Figure~\ref{fig:prec_vs_rec_audio} shows the precision-recall and purity-coverage curves achieved by the two fine-tuned wav2vec2 models when evaluated on the entire audio data. Table~\ref{fig:prec_vs_rec_audio} then lists the numeric results corresponding to the default thresholds.

From the results displayed in Figure~\ref{fig:prec_vs_rec_audio}, it appears that the addition of intermediate boundaries to the training data has had a very minimal effect on the two curves, at least when evaluated only on prosodic boundaries. However, if we look at the false positives, a greater percentage of them now consists of intermediate boundaries as opposed to no-breaks. This could be considered an improvement by itself -- in many use-cases, an intermediate boundary being incorrectly marked as a prosodic boundary is a less problematic mistake than if a location with \emph{no} phrase boundary was marked as such.

\begin{table}[ht]
\centering
\caption{Results on the entire audio files, including boundaries at the ends of sentences, and with a wav2vec2 model fine-tuned a) only on prosodic boundaries (threshold 0.5) or b) also intermediate boundaries (threshold 0.75). \enquote{Pur} and \enquote{Cov} refers to purity and coverage, respectively.}
\label{tab:textresults}
\begin{tabular}{| l | *{6}{S[round-mode=places,table-format=3.2, round-precision=2,detect-weight]} | c c c c |}

\hline
{fine-tuning data} & {Pur} & {Cov} & {Acc} & {P} & {R} & {F1} & {tp} & {fp} & {fn} & {tn} \\

\hline
a) prosodic b. only & 93.8225 & 92.9418 & 94.8698 & 88.223 & 88.9045 & 88.5624 & 1266 & 169 & 158 & 4781 \\

b) pros. \& interm. b. & 92.2868 & 94.3863 & 94.7757 & 90.2583 & 85.8848 & 88.0173 & 1223 & 132 & 201 & 4818 \\

\hline
\end{tabular}
\end{table}

\subsection{Text-based evaluation}
\label{sec:textresults}

In order to compare the results of the wav2vec2 model with the text-based T5 model, we convert the wav2vec2 predictions to the same format -- sequences of words, separated by sentence, with marked prosodic boundaries.

Thus, for this second evaluation, we only consider the predicted phrase boundaries which were matched to word boundaries within the sentence during post-processing. Peaks in the wav2vec2 output which were more than 100\,ms from the nearest end of a word are simply ignored. However, the number of such cases is minimal (3 out of 1435 predicted boundaries at threshold 0.5).

\begin{figure}
    \centering
    \includegraphics[width=0.6\textwidth]{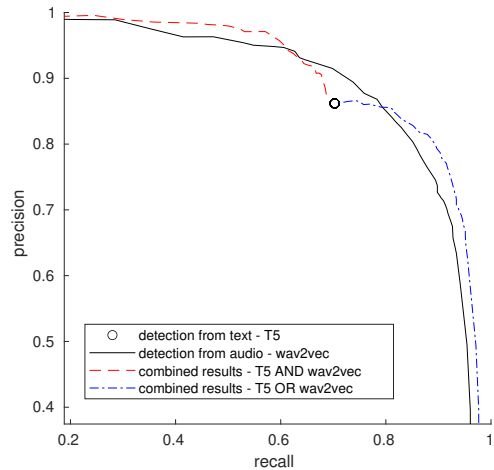}
    \caption{Results evaluated on text - precision and recall of the text-based T5 model, audio-based wav2vec2 model, and their combinations (wav2vec2 fine-tuned only on prosodic boundaries)}
    \label{fig:prec_vs_rec}
\end{figure}

The text-based results are illustrated in Figure~\ref{fig:prec_vs_rec}, which compares the pre\-ci\-sion-re\-call curve of the wav2vec2 model (fine-tuned with prosodic boundaries only) with the results of the T5 model. The latter are shown only as a single point, as there is no threshold to change -- the T5 model directly outputs a sequence of words and prosodic boundaries. 

The graph additionally shows the precision-recall curves for two possible combinations of the two models:

\begin{enumerate}
    \item[a)] prosodic boundaries are marked only where \emph{both} the T5 model and the wav2vec2 model predict them (\enquote{T5 AND wav2vec}),
    \item[b)] prosodic boundaries are marked where \emph{at least one} of the models predicts them (\enquote{T5 OR wav2vec}).
\end{enumerate}

Finally, the numeric results are presented in Tables~\ref{tab:textresults_perspeaker} and~\ref{tab:textresults_summary}: Table~\ref{tab:textresults_perspeaker} lists the individual results of the T5 model and one wav2vec2 model (fine-tuned only on prosodic boundaries) for separate speakers. Table~\ref{tab:textresults_summary} then shows the overall results for both wav2vec2 models and also for the combinations of T5 and wav2vec2.

We can see that in the terms of accuracy and F1, the listed \enquote{T5 OR wav2vec} variants score higher that the individual models alone. However, it is at the cost of slightly reduced precision. Conversely, the \enquote{T5 AND wav2vec} achieve a very high precision of $\sim$94\,\%, but with a relatively low recall of $\sim$60\,\%. Which one of these alternatives is best would depend on the specific application.

\begin{table}[ht]
	\centering
	\caption{Results on individual speakers, wav2vec2 fine-tuned only on prosodic boundaries and with a threshold of 0.5.}
	\label{tab:textresults_perspeaker}
	\begin{tabular}{| l | c | c | *{4}{S[round-mode=places,table-format=3.2, round-precision=2,detect-weight]} | c c c c |}
		
		\hline
		{model} & {speaker} & {\# sent.} & {Acc} & {P} & {R} & {F1} & {tp} & {fp} & {fn} & {tn} \\
		
		\hline 
		T5
		& NRS01 & 36 & 91.4221 & 85.2459 & 64.1975 & 73.2394 & 52 & 9 & 29 & 353 \\ 
		& NRS02 & 60 & 94.2424 & 82.0225 & 76.8421 & 79.3478 & 73 & 16 & 22 & 549 \\ 
		& NRS03 & 38 & 95.7447 & 86.3636 & 83.8235 & 85.0746 & 57 & 9 & 11 & 393 \\ 
		& NRS04 & 31 & 92.757 & 84.9057 & 66.1765 & 74.3802 & 45 & 8 & 23 & 352 \\ 
		& NRS05 & 48 & 92.559 & 87.6712 & 66.6667 & 75.7396 & 64 & 9 & 32 & 446 \\ 
		& NRS06 & 45 & 89.2857 & 92.5373 & 54.386 & 68.5083 & 62 & 5 & 52 & 413 \\ 
		& NRS07 & 33 & 94.6565 & 87.037 & 77.0492 & 81.7391 & 47 & 7 & 14 & 325 \\ 
		& NRS08 & 37 & 93.2755 & 91.0714 & 66.2338 & 76.6917 & 51 & 5 & 26 & 379 \\ 
		& NRS09 & 50 & 94.3978 & 80 & 74.7253 & 77.2727 & 68 & 17 & 23 & 606 \\ 
		& NRS10 & 34 & 94.5736 & 90 & 73.7705 & 81.0811 & 45 & 5 & 16 & 321 \\ 
		& NRS11 & 35 & 94.697 & 86 & 75.4386 & 80.3738 & 43 & 7 & 14 & 332 \\ 
		& NRS12 & 39 & 93.9597 & 85.7143 & 75 & 80 & 54 & 9 & 18 & 366 \\ 
		\cline{2-11}
		& all & 486 & 93.4376 & 86.1799 & 70.2444 & 77.4005 & 661 & 106 & 280 & 4835 \\
		
		\hline
		wav2vec2 
		& NRS01 & 36 & 92.5508 & 81.5789 & 76.5432 & 78.9809 & 62 & 14 & 19 & 348 \\ 
		& NRS02 & 60 & 95.6061 & 82.3529 & 88.4211 & 85.2792 & 84 & 18 & 11 & 547 \\ 
		& NRS03 & 38 & 94.6809 & 77.9221 & 88.2353 & 82.7586 & 60 & 17 & 8 & 385 \\ 
		& NRS04 & 31 & 94.3925 & 84.375 & 79.4118 & 81.8182 & 54 & 10 & 14 & 350 \\ 
		& NRS05 & 48 & 92.0145 & 80.2326 & 71.875 & 75.8242 & 69 & 17 & 27 & 438 \\ 
		& NRS06 & 45 & 96.2406 & 92.7273 & 89.4737 & 91.0714 & 102 & 8 & 12 & 410 \\ 
		& NRS07 & 33 & 95.6743 & 85.4839 & 86.8852 & 86.1789 & 53 & 9 & 8 & 323 \\ 
		& NRS08 & 37 & 93.4924 & 84.058 & 75.3247 & 79.4521 & 58 & 11 & 19 & 373 \\ 
		& NRS09 & 50 & 94.1176 & 76.3441 & 78.022 & 77.1739 & 71 & 22 & 20 & 601 \\ 
		& NRS10 & 34 & 95.0904 & 83.871 & 85.2459 & 84.5528 & 52 & 10 & 9 & 316 \\ 
		& NRS11 & 35 & 95.202 & 76.3889 & 96.4912 & 85.2713 & 55 & 17 & 2 & 322 \\ 
		& NRS12 & 39 & 94.6309 & 82.4324 & 84.7222 & 83.5616 & 61 & 13 & 11 & 362 \\ 
		\cline{2-11}
		& all & 486 & 94.4577 & 82.471 & 82.9968 & 82.7331 & 781 & 166 & 160 & 4775 \\
		\hline
		
	\end{tabular}
\end{table}

\begin{table}[ht]
\centering
\caption{Results on the entire NRS data, using a leave-one-out-approach. \enquote{T5 AND wav2vec2}  places prosodic boundaries only where \emph{both} models predicted them. \enquote{T5 OR wav2vec2} places them where \emph{at least one} of the models did.}
\label{tab:textresults_summary}
\begin{tabular}{| l H H | *{4}{S[round-mode=places,table-format=3.2, round-precision=2,detect-weight]} | c c c c |}

\hline
{model} & {speaker} & {\# sent.} & {Acc} & {P} & {R} & {F1} & {tp} & {fp} & {fn} & {tn} \\
\hline
T5 Model & all & 486 & 93.4376 & 86.1799 & 70.2444 & 77.4005 & 661 & 106 & 280 & 4835 \\
wav2vec2 - f.-t. on pros. b. only & all & 486 & 94.4577 & 82.471 & 82.9968 & 82.7331 & 781 & 166 & 160 & 4775 \\
wav2vec2 - fine-tuned with int. b. & all & 486 & 94.3557 & 85.1211 & 78.4272 & 81.6372 & 738 & 129 & 203 & 4812 \\
\hline
T5 AND wav2vec2 (pros. b.)  & all & 486 & 93.3526 & 93.3754 & 62.9118 & 75.1746 & 592 & 42 & 349 & 4899 \\
T5 OR wav2vec2 (pros. b.) & all & 486 & 94.5427 & 78.7037 & \B 90.3294 & \B 84.1168 & 850 & 230 & 91 & 4711 \\
T5 AND wav2vec2 (with int. b.) & all & 486 & 93.1486 & \B 94.3894 & 60.7864 & 73.9496 & 572 & 34 & 369 & 4907 \\
T5 OR wav2vec2 (with int. b.) & all & 486 & \B 94.6447 & 80.4475 & 87.8852 & 84.002 & 827 & 201 & 114 & 4740 \\

\hline
\end{tabular}
\end{table}

One may also notice that the precision and recall values here are slightly lower than those in Table~\ref{tab:textresults}. This is because of the exclusion of end-of-sentence boundaries. These are naturally much more pronounced in speech, in terms of both intonation and pause, and so the wav2vec2 model can detect them with much greater accuracy than the within-sentence boundaries.

\section{Discussion}

We have shown that the results achieved by the wav2vec2 model surpass those of the text-based T5 model. However, it is important to note that this is still a somewhat \enquote{unfair} comparison:
The locations of phrase breaks are partly subjective and different speakers may place them differently. However, the ground truth labels (provided by phonetic experts) used in our experiments were based on the spoken sentences and therefore likely match the specific phrasing of the speaker.
Thus, the predictions made by the T5 model may not necessarily be \emph{less correct}, they simply do not match the specific speaker.

Another thing to consider is the relatively small amount of data which was available for fine-tuning and testing -- approximately 42 minutes of speech or 486 sentences. Although the wav2vec2 framework is known for being able to achieve good results with small amounts of data, and the results achieved here do indeed look very promising, it is likely that the performance could be improved further if more data were available.

\begin{figure}
    \centering
    \includegraphics[width=0.6\textwidth]{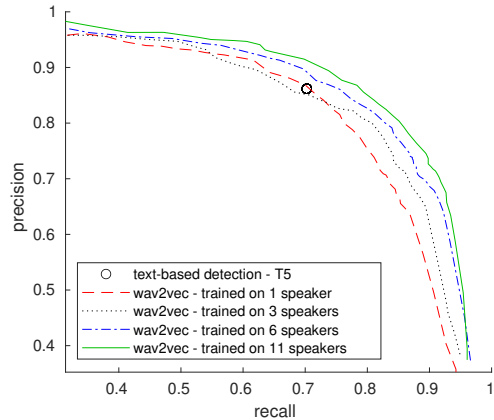}
    \caption{Precision-recall curve for wav2vec2 models fine-tuned using different amounts of training data, evaluated on the text sentences and with the text-based results shown for comparison.}
    \label{fig:trainset_size_comparison}
\end{figure}

This is also suggested by Figure~\ref{fig:trainset_size_comparison}, which compares wav2vec2 models fine-tuned with different amounts of training data: models fine-tuned using only one, three or six of the 12 speakers show a lower precision and recall, indicating that \emph{increasing} the amount of training data could lead to further improvement.

\section{Conclusion and Future Work}

In this paper, we explored the use of the wav2vec 2.0 framework for the detection of prosodic boundaries in speech.

We have found that the relatively straightforward and easy to use wav2vec 2.0 approach works surprisingly well: it does not require text annotation or knowledge of word boundaries (these were only used for evaluation), nor a handcrafted selection of features, yet it achieves very good results, surpassing the text-based T5 model which was used for comparison.

Still, this was, in its essence, only an initial experiment. In the future, we would like to test the approach on a larger amount of more varied data and also explore the possibilities of combining the audio and text modalities within a single model, rather than merely combining the outputs.

\subsubsection{Acknowledgements} 

This research was supported by the Czech Science Foundation (GA CR), project No. GA21-14758S, and by the grant of the University of West Bohemia, project No. SGS-2022-017. 
Computational resources were supplied by the project ``e-Infrastruktura CZ'' (e-INFRA CZ LM2018140) supported by the Ministry of Education, Youth and Sports of the Czech Republic. 

\bibliographystyle{splncs04}
\bibliography{bibliography}
\end{document}